\documentclass{article}%
\usepackage[top=3cm, bottom=3cm, left=3cm, right=3cm]{geometry}
\usepackage{amsmath}
\usepackage{amsfonts}
\usepackage{amssymb}
\usepackage{graphicx}%
\begin{document}

\title{Mirror versus naive assignment in chiral models for the nucleon}
\author{Susanna Gallas and Francesco Giacosa\\\emph{Institut f\"{u}r Theoretische Physik, Johann Wolfgang Goethe -
Universit\"{a}t}\\\emph{Max von Laue--Str. 1, 60438 Frankfurt am Main, Germany}}
\date{\phantom{1}}
\maketitle

\begin{abstract}
In the framework of chiral model(s) based on the linear realization of chiral
symmetry, we investigate the low-energy phenomenological properties of the
nucleon and its chiral partner (identified with either $N(1535)$ or $N(1650)$)
in the naive assignment and compare the results with the ones obtained in the
mirror assignment. We find that, within the naive assignment, we are not able
to reproduce the experimental value of the isospin-odd scattering length,
while the mirror assignment is in good agreement with it for both choices
$N(1535)$ and $N(1650)$. The isospin-even scattering length is not yet
conclusive in either assignment because it depends crucially on the poorly
known scalar mesonic sector. The decay with $\Gamma_{N(1535)\rightarrow N\eta
}$ turns out to be far too small in both the naive and mirror assignments,
while $\Gamma_{N(1650)\rightarrow N\eta}$ is described correctly by both of
them. In conclusions, the mirror assignment with $N(1650)$ as the chiral
partner of the nucleon is the favored configuration.

\end{abstract}

\section{Introduction}

The question of the mass generation of particles is crucial in modern high
energy physics. The standard model predicts the existence of a scalar boson,
the {\emph{Higgs}} particle, which is capable of giving mass to the
fundamental particles by the spontaneous breaking of the electroweak symmetry.
Now, what about the mass of composite objects like the baryons, most notably
the nucleon? The sum of the mass of the three quarks that form a nucleon is
not equal to the mass of the baryon, but is much smaller ($<5\%$). Where does
the dominant contribution to the nucleon mass come from?\newline In this work
we want to shed some light on this question. We focus on two mechanisms in
which we can introduce the nucleon and its chiral partner, denoted as $N$ and
$N^{\ast}$ respectively, by testing the consequences in each case. The field
$N^{\ast}$ is usually identified with the resonance $N(1535),$ but $N(1650)$
is also a viable candidate \cite{PDG}. We study both scenarios in this
work.\newline One way to introduce the chiral partner of the nucleon is via
the $\mathit{naive}$ $\mathit{assignment}$ \cite{Lee,koch,dmitrasinovic}. In
this assignment the mass of the nucleon (and that of its chiral partner) is
generated only through spontaneous chiral symmetry breaking, since a mass term
would not be chirally invariant. (We neglect here the small contribution of
the nonzero bare quark masses.)

The introduction of an explicit mass term, which should occur in a chirally
invariant way, is only possible within the $\mathit{mirror}$
$\mathit{assignment}$. The mirror assignment was analyzed for the first time
in Ref. \cite{Lee} and there dismissed as unusable, as it led to the
unphysical conclusion that the coupling between nucleons and pions vanishes.
This problem was solved later by introducing all terms allowed by chiral
symmetry. The interest in the mirror assignment has been revitalized by the
work in Ref. \cite{Detar:1988kn}, after which it has been investigated in a
variety of works studying the phenomenology of baryons in the vacuum and at
nonzero density \cite{Jido:2001nt,Hosaka:2003hi, Gallas:2009qp,
Zakopane,giuseppe,paeng,sasakimishustin,ziesche,glozman}.

It is still an open question which is the correct assignment in nature. To
this end, both assignments should be investigated and validated by
experimental data such as the axial coupling constants, the decays $N^{\ast
}\rightarrow N\pi$ and $N^{\ast}\rightarrow N\eta$, and the $N\pi$ scattering
lengths: this is the aim of the present work. Namely, while in our previous
study of Ref. \cite{Gallas:2009qp} we concentrated, in the framework of the
so-called extended Linear Sigma Model (eLSM), on the mirror assignment only,
here we construct for the first time the eLSM version also for the naive
assignment. The eLSM contains, besides the usual (pseudo)scalar mesons, also
(axial-)vector degrees of freedom; their presence changes substantially the
properties of baryon phenomenology both in the naive and mirror assignments
allowing for a correct description of the axial coupling constants of $N$ and
$N^{\ast}$. Moreover, in the naive assignment, the absence of (axial-)vector
states implies a complete decoupling of the fields $N$ and $N^{\ast}$
\cite{Jido:2001nt}, thus also implying the (unphysical) vanishing of the decay
$N^{\ast}\rightarrow N\pi.$ The introduction of (axial-)vector d.o.f. implies
a residual $N^{\ast}N\pi$ coupling, thus making also the naive assignment
realistic. However, we shall find that, even in this more complete treatment,
the naive assignment is not capable to reproduce the isospin-odd pion-nucleon
scattering length. We thus arrive at the conclusion that the mirror assignment
is favoured to be the correct way to incorporate baryons in (linear) chiral models.

It should be stressed that our approach is not designed to reach the accuracy
that chiral perturbation theory delivers in the description of pion-nucleon
scattering. Namely, the latter approach is tailor-made to study the
interaction of nucleons with the quasi-Goldstone bosons of low-energy QCD, the
pions, in a systematic framework, see e.g. Refs. \cite{meissner,Baru:2010xn}
and refs. therein. However, our linear chiral approach has the advantage to
incorporate from the very beginning massive resonances (i.e., the chiral
partners), and is thus suited to study their decays, as shown in\ Ref.
\cite{Gallas:2009qp} in the baryonic sector and in Refs.
\cite{denis,Parganlija:2012fy} in the mesonic sector. As a consequence of the
chiral invariance of the Lagrangian, the low-energy theorems are fulfilled.
Moreover, our approach can be easily implemented at nonzero temperature and
density to investigate the chiral phase transition within hadronic matter
\cite{giuseppe,achimtq}. In this context the use of the correct assignment is
mandatory, because it strongly affects the behavior of physical quantities in
a dense and hot medium.

This paper is organized as follows: in section \ref{II} the Lagrangians of
both assignments are studied, in section \ref{III} the results are shown and
finally in section \ref{IV} conclusions and outlooks are presented.\newline
Our units are $\hbar=c=1$, the metric tensor is $g^{\mu\nu}=\mathrm{diag}%
(+,-,-,-)$.

\section{Mirror and naive assignments}

\label{II}

In this section we present the baryonic part of the eLSM in the mirror and
naive assignments, respectively. We concentrate to the two-flavour case,
$N_{f}=2$. The scalar and pseudoscalar fields are included in the matrix
\begin{equation}
\Phi=\sum_{a=0}^{3}\phi_{a}t_{a}=(\sigma+i\eta_{N})\,t^{0}+(\vec{a}_{0}%
+i\vec{\pi})\cdot\vec{t}\;, \label{scalars}%
\end{equation}
where $\vec{t}=\vec{\tau}/2,$ with the vector of Pauli matrices $\vec{\tau}$,
and $t^{0}=\mathbf{1}_{2}/2$. Under $U(N_{f}=2)_{R}\times U(N_{f}=2)_{L}$
chiral symmetry, $\Phi$ transforms as $\Phi\rightarrow U_{L}\Phi
U_{R}^{\dagger}$. The vector and axial-vector fields are contained in the
matrices
\begin{subequations}
\label{vectors}%
\begin{align}
V^{\mu}  &  =\sum_{a=0}^{3}V_{a}^{\mu}t_{a}=\omega^{\mu}\,t^{0}+\vec{\rho
}^{\mu}\cdot\vec{t}\;,\\
A^{\mu}  &  =\sum_{a=0}^{3}A_{a}^{\mu}t_{a}=f_{1}^{\mu}\,t^{0}+\vec{a_{1}%
}^{\mu}\cdot\vec{t}\;.
\end{align}
From these fields, we define the right- and left-handed vector fields as
$R^{\mu}\equiv V^{\mu}-A^{\mu}$and $L^{\mu}\equiv V^{\mu}+A^{\mu}$. Under
$U(2)_{R}\times U(2)_{L}$ these fields change as $R^{\mu}\rightarrow
U_{R}R^{\mu}U_{R}^{\dagger}\,,\;L^{\mu}\rightarrow U_{L}L^{\mu}U_{L}^{\dagger
}$. The identification of mesons with particles listed in Ref.\ \cite{PDG} is
as follows: the fields $\vec{\pi}$ and $\eta_{N}$ correspond to the pion and
the $SU(2)$ counterpart of the $\eta$ meson, $\eta_{N}\equiv(\overline
{u}u+\overline{d}d)/\sqrt{2}$, with a mass of about $700$ MeV. This value can
be obtained by unmixing the physical $\eta$ and $\eta^{\prime}$ mesons, which
also contain $\overline{s}s$ contributions. The fields $\omega^{\mu}$ and
$\vec{\rho}^{\mu}$ represent the vector mesons $\omega(782)$ and $\rho(770)$,
and the fields $f_{1}^{\mu}$ and $\vec{a_{1}}^{\mu}$ represent the
axial-vector mesons $f_{1}(1285)$ and $a_{1}(1260)$, respectively. \newline
The identification of the $\sigma$ and $\vec{a}_{0}$ fields is controversial,
the possible pairs are $\{f_{0}(500),a_{0}(980)\}$ and $\{f_{0}(1370),a_{0}%
(1450)\}$, the latter one being in agreement with the phenomenology, see Refs.
\cite{denis,Parganlija:2012fy} in which the mesonic part of the Lagrangian is
presented in depth. Some conclusions are however essential for the following
studies of the baryonic part as well: the chiral condensate $\sigma
_{0}=\left\langle 0\left\vert \sigma\right\vert 0\right\rangle =Zf_{\pi}$
emerges upon spontaneous chiral symmetry breaking in the mesonic sector. The
parameter $f_{\pi}=92.2$ MeV is the pion decay constant and $Z=1.81$ is the
wave-function renormalization constant of the pseudoscalar fields (see the
determination of $Z$ in Sec. 3).

Turning to the baryon sector, we have the baryon doublets $\Psi_{1}$ and
$\Psi_{2}$, where $\Psi_{1}$ has positive parity and $\Psi_{2}$ negative
parity. The physical fields $N$ and $N^{\ast}$ are related to the spinors
$\Psi_{1}$ and $\Psi_{2}$ through:
\end{subequations}
\begin{align}
\Psi_{1}  &  =\frac{1}{\sqrt{2\cosh\delta}}(Ne^{\delta/2}+\gamma_{5}N^{\ast
}e^{-\delta/2})\;,\nonumber\\
\Psi_{2}  &  =\frac{1}{\sqrt{2\cosh\delta}}(\gamma_{5}Ne^{-\delta/2}-N^{\ast
}e^{\delta/2})\;. \label{rotationfelder}%
\end{align}
The quantity $\delta$ parametrizes the mixing between $\Psi_{1}$ and $\Psi
_{2}$: for $\delta\rightarrow\infty$ the mixing of the fields disappears,
implying $\Psi_{1}=N$ and $\Psi_{2}=N^{\ast}$; for $\delta=0$ the mixing is
maximal. \newline The fundamental difference between the mirror assignment and
the naive assignment is the transformation of these baryon doublets under the
chiral group $U(2)_{R}\times U(2)_{L}$, which we discuss in the following.
\newline

\begin{enumerate}
\item \textit{Mirror assignment\newline}

The nucleons $N$ and $N^{\ast}$ belong to the same multiplet and thus are true
chiral partners \cite{Jido:2001nt, Nemoto:1998um}. In this assignment the
left-handed and the right-handed parts of $\Psi_{1}$ and $\Psi_{2}$ ($\Psi
_{i}=\Psi_{i,R}+\Psi_{i,L}$, $i=1,2$) transform as:
\begin{align}
&  \Psi_{1R}\rightarrow U_{R}\Psi_{1R}\,,\;\Psi_{1L}\rightarrow U_{L}\Psi
_{1L}\,,\;\nonumber\\
&  \Psi_{2R}\rightarrow U_{L}\Psi_{2R}\,,\;\Psi_{2L}\rightarrow U_{R}\Psi
_{2L}\;, \label{mirror}%
\end{align}
i.e. $\Psi_{2}$ transforms in a \textquotedblleft mirror way\textquotedblright%
\ under chiral transformation. These field transformations allow to write down
a baryonic Lagrangian with a chirally invariant mass term for the fermions,
which is parametrized by $m_{0}$. The full Lagrangian of the eLSM (for
$N_{f}=2$) in the baryonic sector reads \cite{Gallas:2009qp}:
\begin{align}
\mathcal{L}_{\mathrm{mirror}}  &  =\overline{\Psi}_{1L}i\gamma_{\mu}%
D_{1L}^{\mu}\Psi_{1L}+\overline{\Psi}_{1R}i\gamma_{\mu}D_{1R}^{\mu}\Psi
_{1R}+\overline{\Psi}_{2L}i\gamma_{\mu}D_{2R}^{\mu}\Psi_{2L}+\overline{\Psi
}_{2R}i\gamma_{\mu}D_{2L}^{\mu}\Psi_{2R}\nonumber\\
&  -\widehat{g}_{1}\left(  \overline{\Psi}_{1L}\Phi\Psi_{1R}+\overline{\Psi
}_{1R}\Phi^{\dagger}\Psi_{1L}\right)  -\widehat{g}_{2}\left(  \overline{\Psi
}_{2L}\Phi^{\dagger}\Psi_{2R}+\overline{\Psi}_{2R}\Phi\Psi_{2L}\right)
\nonumber\\
&  -m_{0}(\overline{\Psi}_{1L}\Psi_{2R}-\overline{\Psi}_{1R}\Psi
_{2L}-\overline{\Psi}_{2L}\Psi_{1R}+\overline{\Psi}_{2R}\Psi_{1L})\;\text{,}
\label{nucl lagra}%
\end{align}
where $D_{1R}^{\mu}=\partial^{\mu}-ic_{1}R^{\mu}$, $D_{1L}^{\mu}=\partial
^{\mu}-ic_{1}L^{\mu}$, and $D_{2R}^{\mu}=\partial^{\mu}-ic_{2}R^{\mu}$,
$D_{2L}^{\mu}=\partial^{\mu}-ic_{2}L^{\mu}$ are the covariant derivatives for
the nucleonic fields, with the dimensionless coupling constants $c_{1}$ and
$c_{2}$. The interaction of the baryonic fields with the scalar and
pseudoscalar mesons is parametrized by the dimensionless coupling constants
$\widehat{g}_{1}$ and $\widehat{g}_{2}$. The last term in Eq.
(\ref{nucl lagra})%
\begin{equation}
\mathcal{L}_{m_{0}}=-m_{0}(\overline{\Psi}_{1L}\Psi_{2R}-\overline{\Psi}%
_{1R}\Psi_{2L}-\overline{\Psi}_{2L}\Psi_{1R}+\overline{\Psi}_{2R}\Psi_{1L})\;
\label{lm0}%
\end{equation}
is the chirally invariant mass term for the baryons. The masses of the nucleon
and its chiral partner are obtained by diagonalizing the corresponding mass
matrix in the Lagrangian:
\begin{align}
&  m_{N}=\sqrt{\left[  \frac{\widehat{g}_{1}+\widehat{g}_{2}}{4}\right]
^{2}\sigma_{0}^{2}+m_{0}^{2}}+\frac{\widehat{g}_{1}-\widehat{g}_{2}}{4}%
\sigma_{0}\;\text{,}\nonumber\\
&  m_{N^{\ast}}=\sqrt{\left[  \frac{\widehat{g}_{1}+\widehat{g}_{2}}%
{4}\right]  ^{2}\sigma_{0}^{2}+m_{0}^{2}}-\frac{\widehat{g}_{1}-\widehat
{g}_{2}}{4}\sigma_{0}\;\text{.} \label{nuclmasses}%
\end{align}
The parameter $\delta$ entering in Eqs. (\ref{rotationfelder}) reads:
\begin{equation}
\mathrm{sinh}\;\delta=\frac{(\widehat{g}_{1}+\widehat{g}_{2})\sigma_{0}%
}{4m_{0}}\;\text{.}%
\end{equation}
From these expressions it is evident that the baryonic masses are generated by
chiral symmetry breaking trough the appearance of the chiral condensate
$\sigma_{0}$ and by the mass parameter $m_{0}$. In particular, the chiral
condensate originates a splitting of the masses in the chirally broken phase.
In the limit $\sigma_{0}\rightarrow0$ both masses become degenerate, but do
not vanish: $m_{N}=m_{N^{\ast}}=m_{0}\neq0$ (see Fig. 1, where the masses are
plotted as function of $\sigma_{0}$). The numerical value of $m_{0}$ is
determined through a fit procedure, see Sec. 3 and Ref. \cite{Gallas:2009qp}
for details. In the case $N^{\ast}\equiv N(1535)$ one obtains $m_{0}%
\simeq500\;\mathrm{MeV}$, while for $N^{\ast}\equiv N(1650)$ one obtains a
even larger value $m_{0}\simeq700\;\mathrm{MeV}$. These results mean that
$m_{0}$ is sizable and is therefore an important component of the baryonic
masses. These values are larger than the one originally found in the
pioneering work of Ref. \cite{Detar:1988kn}, in which $m_{0}\simeq
200\;\mathrm{MeV}$ was found.

The expressions for the axial coupling constants of the nucleon and its chiral
partner are given by \cite{Gallas:2009qp}:
\begin{equation}
g_{A}^{N}=\frac{1}{2\cosh\delta}\left(  g_{A}^{(1)}\,e^{\delta}+g_{A}%
^{(2)}\,e^{-\delta}\right)  \;,\;\;g_{A}^{N^{\ast}}=\frac{1}{2\cosh\delta
}\left(  g_{A}^{(1)}\,e^{-\delta}+g_{A}^{(2)}\,e^{\delta}\right)  \;,
\label{axialcoupl}%
\end{equation}
where
\begin{equation}
g_{A}^{(1)}=1-\frac{c_{1}}{g_{1}}\left(  1-\frac{1}{Z^{2}}\right)
,\;\;g_{A}^{(2)}=-1+\frac{c_{2}}{g_{1}}\left(  1-\frac{1}{Z^{2}}\right)
\text{ ,} \label{gas}%
\end{equation}

where the parameter $g_{1}=6.1$ describes the interaction of (axial-)vector
mesons with (pseudo)scalar mesons \cite{denis}. It is evident that, due to the
presence of the parameter $c_{1}$ and $c_{2}$ which arise from the interaction
terms of the baryons with (axial-)vector degrees of freedom, the
axial-coupling constant $g_{A}^{N}$ can be larger than unity. (The limitation
$\left\vert g_{A}^{N}\right\vert \leq1$ holds if vector mesons are not
included \cite{dmitra2}). This shows that the inclusion of the (axial-)vector
states is important for the correct description of the phenomenology, see the
results in the next section.

Note, the Lagrangian of Eq. (\ref{nucl lagra}) is invariant under chiral
transformation. This means that in the present version of the model the
explicit breaking of chiral symmetry is confined to the mesonic sector, in
which the pions acquire a small but nonzero mass. Indeed, it would not be
difficult to include an explicit symmetry breaking mass term also in the
baryonic sector, but in view of the purposes of this work and in order to keep
the number of parameters as small as possible, we leave this task for the
future. In particular, such an explicit symmetry breaking will be unavoidable
when the three-flavor version of the model will be studied. However, it is
important to stress that even if no explicit symmetry breaking term in the
baryonic Lagrangian is present the so-called pion-nucleon sigma term
$\sigma_{\pi N}$ (see for instance Refs.
\cite{dmitrasinovic,latticesigmapion,chaos,jorge} and refs. therein), does not
vanish \cite{koch}. In fact, this term, which describes the contribution to
the nucleon mass which arises from the explicit symmetry breaking, acquires a
nonzero (and quite sizable) contribution from the explicit symmetry breaking
encoded in the chiral condensate $\sigma_{0}$. The numerical results for
$\sigma_{\pi N}$ are discussed in the next session. \textbf{ }%
%TCIMACRO{\FRAME{ftbpFU}{3.8251in}{2.3142in}{0pt}{\Qcb{The masses of the
%baryons $N$ and $N^{\ast}$ in the a) mirror assignment (Eq. (\ref{nuclmasses}%
%), parameters from Eq. (\ref{mirrorparam1})) and b) in the naive assignment
%(Eq. (\ref{mN_ghut}), parameters from Eq. (\ref{naiveparam})) as a function of
%the chiral condensate $\sigma_{0}$: a) for vanishing $\sigma_{0}$ the masses
%acquire the same non-zero value $m_{N}=m_{N^{\ast}}=m_{0}$, while in b) they
%vanish in this limit. On the very right of the plot, the masses reach their
%physical values, $m_{N}=939$ MeV and $m_{N^{\ast}}=1535$ MeV, in both
%assignments.}}{}{masses.eps}{\special{ language "Scientific Word";
%type "GRAPHIC";  maintain-aspect-ratio TRUE;  display "USEDEF";
%valid_file "F";  width 3.8251in;  height 2.3142in;  depth 0pt;
%original-width 6.4991in;  original-height 3.9202in;  cropleft "0";
%filename 'masses.eps';file-properties "NPEU";}}}%
%BeginExpansion
\begin{figure}
[ptb]
\begin{center}
\includegraphics[
height=2.3142in,
width=3.8251in
]%
{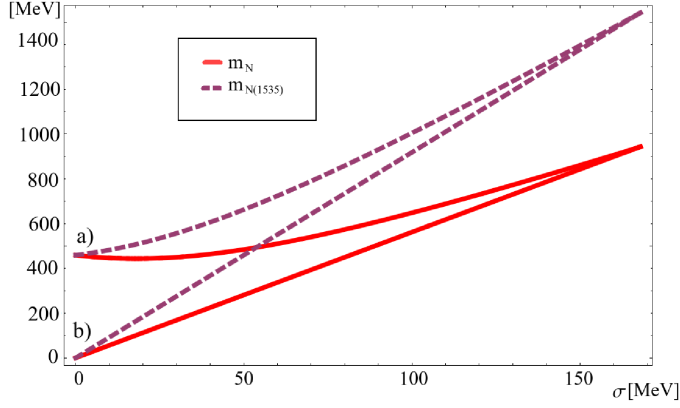}%
\caption{The masses of the baryons $N$ and $N^{\ast}$ in the a) mirror
assignment (Eq. (\ref{nuclmasses}), parameters from Eq. (\ref{mirrorparam1}))
and b) in the naive assignment (Eq. (\ref{mN_ghut}), parameters from Eq.
(\ref{naiveparam})) as a function of the chiral condensate $\sigma_{0}$: a)
for vanishing $\sigma_{0}$ the masses acquire the same non-zero value
$m_{N}=m_{N^{\ast}}=m_{0}$, while in b) they vanish in this limit. On the very
right of the plot, the masses reach their physical values, $m_{N}=939$ MeV and
$m_{N^{\ast}}=1535$ MeV, in both assignments.}%
\end{center}
\end{figure}
%EndExpansion

\item \textit{ Naive assignment\newline}

In this assignment both the left-handed and the right-handed part of the
states $\Psi_{1}$ and $\Psi_{2}$ behave in the same way under chiral
transformation:
\begin{align}
&  \Psi_{1R}\rightarrow U_{R}\Psi_{1R}\;,\;\Psi_{1L}\rightarrow U_{L}\Psi
_{1L}\,,\;\nonumber\\
&  \Psi_{2R}\rightarrow U_{R}\Psi_{2R}\,,\;\Psi_{2L}\rightarrow U_{L}\Psi
_{2L}\;. \label{naivetrans}%
\end{align}
By introducing the chiral partner in this way, a nucleon mass term $\sim
m_{N}\bar{\Psi}\Psi$ is not allowed because it explicitly breaks chiral
symmetry. The Lagrangian in the naive assignment reads:
\begin{align}
\mathcal{L}_{\mathrm{naive}}  &  =\overline{\Psi}_{1L}i\gamma_{\mu}D_{1L}%
^{\mu}\Psi_{1L}+\overline{\Psi}_{1R}i\gamma_{\mu}D_{1R}^{\mu}\Psi
_{1R}+\overline{\Psi}_{2L}i\gamma_{\mu}D_{2L}^{\mu}\Psi_{2L}+\overline{\Psi
}_{2R}i\gamma_{\mu}D_{2R}^{\mu}\Psi_{2R}\nonumber\\
&  +c_{12}\overline{\Psi}_{1R}\gamma_{\mu}R^{\mu}\Psi_{2R}+c_{12}%
\overline{\Psi}_{2R}\gamma_{\mu}R^{\mu}\Psi_{1R}-c_{12}\overline{\Psi}%
_{1L}\gamma_{\mu}L^{\mu}\Psi_{2L}-c_{12}\overline{\Psi}_{2L}\gamma_{\mu}%
L^{\mu}\Psi_{1L}\nonumber\\
&  -\widehat{g}_{1}\left(  \overline{\Psi}_{1L}\Phi\Psi_{1R}+\overline{\Psi
}_{1R}\Phi^{\dagger}\Psi_{1L}\right)  -\widehat{g}_{2}\left(  \overline{\Psi
}_{2L}\Phi\Psi_{2R}+\overline{\Psi}_{2R}\Phi^{\dagger}\Psi_{2L}\right)
\nonumber\\
&  +\widehat{g}_{12}\overline{\Psi}_{1R}\Phi^{\dagger}\Psi_{2L}-\widehat
{g}_{12}\overline{\Psi}_{1L}\Phi\Psi_{2R}+\widehat{g}_{12}\overline{\Psi}%
_{2L}\Phi\Psi_{1R}-\widehat{g}_{12}\overline{\Psi}_{2R}\Phi^{\dagger}\Psi
_{1L}\;. \label{Lagnaive}%
\end{align}
Note that, indeed, some terms not permitted in the mirror assignment turn out
to be chirally invariant in the naive case. They are parametrized by the
coupling constants $c_{12}$ and $\widehat{g}_{12}$. The spinors $\Psi_{1}$ and
$\Psi_{2}$ are replaced by the physical fields as dictated by eq.
(\ref{rotationfelder}), where now the mixing parameter $\delta$ is given by:
\begin{equation}
\mathrm{sinh}\;\delta=\frac{\widehat{g}_{1}+\widehat{g}_{2}}{2\widehat{g}%
_{12}}\;. \label{misch}%
\end{equation}

There are however crucial differences between the term proportional to
$\widehat{g}_{12}$ in Eq. (\ref{Lagnaive}) w.r.t. the term proportional to
$m_{0}$ in the mirror case, Eq. (\ref{nucl lagra}). First, the parameter
$\widehat{g}_{12}$ is dimensionless, while $m_{0}$ has dimension of energy.
Second, the $\widehat{g}_{12}$-term involves the same interactions as the
terms proportional to $\widehat{g}_{1}$ and $\widehat{g}_{2}$ and therefore
can be completely eliminated by a suitable field transformation. (Such an
operation is not possible in the mirror case, where the elimination of the
mixing for the quadratic mass terms does not imply an elimination of the
effects of the mixing in the interaction terms). Then, in the naive
assignment, by expressing the Lagrangian in terms of the physical fields $N$
and $N^{\ast}$, we obtain the following form in which only terms which couple
the nucleon, its chiral partner and the pion are included:
\begin{align}
\mathcal{L}_{\mathrm{naive}}  &  =\overline{N}i\gamma_{\mu}\partial^{\mu
}N+\overline{N}^{\ast}i\gamma_{\mu}\partial^{\mu}N^{\ast}-\tilde{\widehat{g}%
}_{1}\overline{N}(\sigma+Zi\gamma_{5}\vec{\pi}\cdot\vec{\tau})N-\tilde
{\widehat{g}}_{2}\overline{N}^{\ast}(\sigma+Zi\gamma_{5}\vec{\pi}\cdot
\vec{\tau})N^{\ast}\nonumber\\
&  -Zw\tilde{c}_{1}\overline{N}\gamma_{5}\gamma^{\mu}\partial_{\mu}\vec{\pi
}\cdot\vec{\tau}N-Zw\tilde{c}_{2}\overline{N}^{\ast}\gamma_{5}\gamma^{\mu
}\partial_{\mu}\vec{\pi}\cdot\vec{\tau}N^{\ast}\nonumber\\
&  -Zw\tilde{c}_{12}\left(  \overline{N}\gamma^{\mu}\partial_{\mu}\vec{\pi
}\cdot\vec{\tau}N^{\ast}+\overline{N}^{\ast}\gamma^{\mu}\partial_{\mu}\vec
{\pi}\cdot\vec{\tau}N\right)  \;+... \label{Lagnaive2}%
\end{align}
The coupling constants with \textquotedblleft tilde\textquotedblright\ are
combinations of the original couplings, $\widehat{g}_{1},\;\widehat{g}%
_{2},\;\widehat{g}_{12},\;c_{1},$ $\;c_{2},\;c_{12}$ and $\delta$, reducing in
this way the number of parameters from six to five (see Appendix
\ref{appendixB}): $\tilde{\widehat{g}}_{1},\;\tilde{\widehat{g}}_{2}%
,\;\tilde{c}_{1},\;\tilde{c}_{2}$ and $\tilde{c}_{12}$. The masses of $N$ and
$N^{\ast}$ can be easily obtained from Eq. (\ref{Lagnaive2}) (see Fig. 1 for a
plot as function of $\sigma_{0}$):
\begin{equation}
m_{N}=\tilde{\widehat{g}}_{1}\sigma_{0}\;\;\;\;\mathrm{and}\;\;\;\;m_{N^{\ast
}}=\tilde{\widehat{g}}_{2}\sigma_{0}\;. \label{mN_ghut}%
\end{equation}

In terms of $N$ and $N^{\ast}$ there are no interactions of the type
$\overline{N}i\gamma_{5}\vec{\pi}\cdot\vec{\tau}N^{\ast}$. Therefore the
coupling between the nucleon, its chiral partner and the pion takes the simple
form $\overline{N}^{\ast}\gamma_{5}\gamma^{\mu}\partial_{\mu}\vec{\pi}%
\cdot\vec{\tau}N+$ $h.c.$, which stems solely from the shift of the
axial-vector meson $\vec{a}_{1}\rightarrow\vec{a}_{1}+Zw\partial_{\mu}\vec
{\pi}$, which guarantees the disappearance of non-diagonal terms in the
mesonic sector of the theory \cite{denis,Parganlija:2012fy,gasioro}.
Explicitly, the parameter $w$ reads:
\begin{equation}
w=\frac{g_{1}\sigma_{0}}{m_{a_{1}}^{2}}\text{ ,} \label{w}%
\end{equation}
where $m_{a_{1}}=1230$ MeV is the mass and the already introduced parameter
$g_{1}=6.1$ describes the interaction of (axial-)vector mesons with
(pseudo)scalar ones. As a consequence, the quantity $w$ takes the numerical
value $w=67\cdot10^{-5}$ MeV$^{-1}$.

In a scenario in which (axial-)vector mesons decouple, the parameter $g_{1}$
vanishes, and so does the parameter $w$ in Eq. (\ref{w}). Therefore the
nucleon and its chiral partner decouple completely \cite{Jido:2001nt} and
decay processes such as $N^{\ast}\rightarrow N\pi$ are not be possible
\cite{Jido:2001nt}. Thus, the naive assignment without (axial-)vector mesons
is explicitly unphysical and can be immediately discarded, because the decay
rate $N^{\ast}\rightarrow N\pi$ has been measured to be nonzero for both
candidates $N(1535)$ and $N(1650)$. With the inclusion of (axial-)vector
d.o.f. this evident drawback is eliminated but, as we shall show, the
isospin-odd scattering length turns out to be anyhow poorly described in the
naive assignment.

The axial coupling constants of the nucleon and its chiral partner are given
by:
\begin{equation}
g_{A}^{(N)}=1-2Zwf_{\pi}\tilde{c}_{1}\;\;,\;\;\;g_{A}^{(N^{\ast})}%
=1-2Zwf_{\pi}\tilde{c}_{2}\;\text{,}%
\end{equation}
where $w$ is given by Eq. (\ref{w}). It is also here visible that the axial
coupling constant $g_{A}^{(N)}$ is not limited by unity due to the interaction
with (axial-)vector states. Moreover, just as in the mirror case, the
Lagrangian (\ref{Lagnaive}) does not include terms with explicit breaking of
chiral symmetry, but the pion-nucleon sigma term $\sigma_{\pi N}$ does not
vanish, see Sec. 3.2 for a numerical evaluation.
\end{enumerate}

\section{Results}

\label{III}

\subsection{Mirror assignment}

We first review the case in which $N^{\ast}$ is identified with the resonance
$N(1535)$. We perform a fit of the four parameters $m_{0},$ $c_{1},$ $c_{2},$
and $Z$ by using the following five experimental and lattice quantities: (i)
The axial-coupling constant of the nucleon: $g_{A}^{(N,exp)}=1.267\pm0.004$
\cite{PDG}. (ii) The lattice result for the axial-coupling constant of
$N^{\ast}(1535)$: $g_{A}^{(N^{\ast})}=0.2\pm0.3$ \cite{Takahashi}. (iii) The
decay width of $N(1535)$ into $N\pi$: $\Gamma_{N^{\ast}\rightarrow N\pi
}=67.5\pm11.2$ MeV \cite{PDG}. (iv) The electromagnetic decay of the
$a_{1}(1230)$ meson into $\pi\gamma$: $\Gamma_{a_{1}\rightarrow\gamma\pi
}=(0.640\pm0.250)$ MeV \cite{PDG}. (v) The strong decay of the $a_{1}(1230)$
meson into $\rho\pi$: $\Gamma_{a_{1}\rightarrow\rho\pi}=(425\pm175)$ MeV. This
value is obtained by using the range $250$-$600$ MeV quoted by \cite{PDG},
under the assumption that this decay channel is dominant. This is in well
agreement with the detailed experimental analysis of Ref. \cite{Asner:1999kj}
and with the recent phenomenological study of the nonet of axial-vector mesons
in Ref. \cite{florianlisa}. The results read ($\chi_{\text{min}}^{2}=0.64$):
\begin{equation}
m_{0}=459\pm117\text{ MeV},\text{ }c_{1}=-2.65\pm0.18\text{ , }c_{2}%
=10.2\pm2.6\text{ , }Z=1.81\pm0.07\text{ .} \label{mirrorparam1}%
\end{equation}
With this value of $m_{0},$ the parameters $\widehat{g}_{1}$ and $\widehat
{g}_{2}$ are: $\widehat{g}_{1}=10.2\pm0.7$ , $\widehat{g}_{2}=17.3\pm0.8.$
Notice that the performed fit is similar but not equal to the one of Ref.
\cite{Gallas:2009qp}: the inclusion of point (v) was not present in the fit of
that work. As a consequence, the present values are slightly different from
the ones of Ref. \cite{Gallas:2009qp}, but it turns out that the overall
phenomenology does not change much (for instance, $m_{0}$ was determined to be
$460\pm136$ MeV). The biggest change is the increased (and less uncertain)
value of the parameter $Z$, which is now in good agreement with the
tree-flavor determination of Ref. \cite{Parganlija:2012fy}.

Having determined the parameters, it is possible to calculate the isospin-odd
scattering length:%
\begin{equation}
a_{0}^{(-)}=(6.41\pm0.17)\cdot10^{-4}\text{ }\mathrm{MeV}^{-1}\;, \label{a0m}%
\end{equation}
which is in very well agreement with the experimental value $a_{0,exp}%
^{(-)}=(6.4\pm0.1)\cdot10^{-4}$ measured at the Paul Scherrer Institute (PSI,
Switzerland) with pionic hydrogen and deuterium X-ray experiments
\cite{gotta}. It is also interesting to compare our result with the one
presented in Ref. \cite{Baru:2010xn}, where the calculation of pion-nucleon
scattering lengths was performed by using chiral perturbation theory (ChPT)
including isospin-violating corrections: $a_{0,ChPT}^{(-)}=(6.16\pm
0.06)\cdot10^{-4}\;$ MeV$^{-1}$. Thus, our value is compatible with the ChPT
result as well. Note, in our previous evaluation of Ref. \cite{Gallas:2009qp}
the scattering length had the value $a_{0}^{(-)}=(6.04\pm0.63)\cdot10^{-4}$
MeV$^{-1}$, which is compatible with the result of Eq. (\ref{a0m}).
Interestingly, the new result has a smaller error in virtue of the improved fit.

The isospin-even scattering length $a_{0}^{(+)}$ depends on the mass of the
scalar-isoscalar $\sigma$. The $\sigma$-meson cannot be assigned unambiguously
to a specific scalar listed in \cite{PDG}, see Refs.
\cite{denis,Parganlija:2012fy,scalars,tetraquark} and refs. therein.
Therefore, the isoscalar scattering length is plotted in Fig. 2 (left panel)
for values of $m_{\sigma}$ varying between $400$ and $1300$ MeV. An agreement
with the experimental (or even ChPT) band is possible for values of
$m_{\sigma}$ smaller than $500$ MeV. While such a low mass is not favoured by
recent studies on light scalar states (the assignment $\sigma\equiv
f_{0}(1370)$ is the favored one), the result shows that a low-scalar field,
identified by the resonance $f_{0}(500)$ and interpreted as a tetraquark or a
molecular state \cite{scalars,tetraquark}, is needed.\ Although not present
here, it can be easily coupled to the model, see Ref.
\cite{Gallas:2009qp,giuseppe} and the discussions in Sec. 4.

A clear problem of the scenario with $N^{\ast}(1535)$ as the chiral partner is
the theoretical result for the decay width $\Gamma_{N^{\ast}\rightarrow N\eta
}^{theor}=4.9\pm0.8$ MeV, which should be compared with the experimental value
$\Gamma_{N^{\ast}\rightarrow N\eta}^{exp}=63\pm10.5$ MeV. Due to the mismatch
in the decay channel $N^{\ast}\rightarrow N\eta,$ one may wonder if the chiral
partner of the nucleon is the resonance $N^{\ast}\equiv N(1650)$.\textbf{ }

For the case $N^{\ast}\equiv N(1650)$ we follow the same strategy, but the
numerical values for the axial coupling constant and the decay width refer now
to the resonance $N(1650)$: $g_{A}^{(N^{\ast})}=0.55\pm0.2$ \cite{Takahashi},
$\Gamma_{N^{\ast}\rightarrow N\pi}^{exp}=105\pm21$ MeV \cite{PDG}. The
parameters $m_{0}$, $c_{1},$ $c_{2},\widehat{g}$, and $Z$ read ($\chi
_{\text{min}}^{2}=0.64$)
\begin{equation}
m_{0}=659\pm146\text{ MeV}\text{ , }c_{1}=-2.80\pm0.21\text{ , }c_{2}%
=13.2\pm0.8\text{ , }Z=1.81\pm0.07\text{ .}%
\end{equation}
The corresponding values of $\widehat{g}_{1}$ and $\widehat{g}_{2}$ are:
$\widehat{g}_{1}=9.1\pm1.1$ , $\widehat{g}_{2}=17.6\pm1.2.$

The result for the isospin-odd scattering length is similar to the previous
scenario:
\begin{equation}
a_{0}^{(-)}=(6.25\pm0.19)\cdot10^{-4}\text{ MeV}^{-1}\;,
\end{equation}
which is in well agreement with the experimental as well as the ChPT values
quoted above. (Also in this case, the theoretical error is sizably smaller
than in Ref. \cite{Gallas:2009qp}, where we had $a_{0}^{(-)}=(5.90\pm
0.46)\cdot10^{-4}$ MeV$^{-1}$).

For the isospin-even scattering length a similar discussion holds, see Fig. 2
(right panel). Interestingly, there is no issue here with the decay into the
$\eta$ meson: the theoretical decay width $\Gamma_{N^{\ast}\rightarrow N\eta
}^{theor}=12.5\pm2.5$ MeV agrees well with the experimental value
$\Gamma_{N^{\ast}\rightarrow N\eta}^{exp}=15.0\pm3.0$ MeV, thus rendering this
scenario favoured.

We summarize the situation of the mirror assignment in Table 1 and\ Fig. 2;
all the relevant experimental values are in Table 2.

As a last step, we turn to the discussion of the pion-nucleon sigma term,
which in the present model with no explicit breaking term in the baryonic
sector takes the form%
\begin{equation}
\sigma_{\pi N}=m_{N}-\left(  m_{N}\right)  _{\sigma_{0}\rightarrow
\sigma_{0,\text{CL}}}\text{ ,}%
\end{equation}
where $\sigma_{0,\text{CL}}$ is the value of the chiral condensate in the
chiral limit. Thus, $\sigma_{\pi N}$ represents the mass contribution which
arises from the explicit symmetry breaking. Using the numerical values of Ref.
\cite{Parganlija:2012fy} we obtain $\sigma_{0,\text{CL}}=164$ MeV, out of
which%
\begin{equation}
\sigma_{\pi N}\simeq15\text{ MeV.}%
\end{equation}
This value is slightly smaller than the recent lattice value $\sigma_{\pi
N}=37\pm8\pm6$ \cite{latticesigmapion} and the theoretical evaluation based on
the CHAOS data $\sigma_{\pi N}=44\pm12$ \cite{chaos}. A somewhat larger value
$\sigma_{\pi N}=59\pm7$ MeV has been found by using chiral perturbation in
Ref. \cite{jorge}. It is then clear that our value, although smaller then what
is found in other approaches, is non-negligible and shows that the explicit
symmetry breaking encoded in the chiral condensate has an important effect on
this quantity. The `missing part' of $\sigma_{\pi N}$ could be easily
accounted by including a standard mass-term for the nucleon fields. This step
will also be crucial (together with the proper inclusion of all light scalar
fields, see conclusions) for a more accurate determination of the isospin-even
scattering length $a_{0}^{(+)}.$ In fact, chiral perturbation theory shows
that the quantities $\sigma_{\pi N}$ and $g_{A}$ directly enter in the lowest
order expression of $a_{0}^{(+)}$ and that the corresponding terms have the
opposite sign, see e.g. Ref. \cite{weisebuch}. This is not the case for the
isospin-odd scattering length, which in lowest-order chiral perturbation
theory depends only on the nucleon mass, the pion mass, and the pion decay
constant: $a_{0\text{,LO-ChPT}}^{(-)}=(1+m_{\pi}/m_{N})\frac{m_{\pi}}{8\pi
f_{\pi}^{2}}\simeq5.69\cdot10^{-4}$ [MeV$^{-1}$]. This value is however about
$15\%$ too small when compared to the experimental result. In the framework of
ChPT the gap is filled by NLO corrections \cite{weisebuch}, while in our
approach the presence of additional mesonic and baryonic resonances allows to
obtain the correct result. (The correspondence between the two approaches can
be formally obtained by integrating out heavier resonances in our eLSM. When
the latter are infinitely heavy, one obtains ChPT in lowest order).

\bigskip

\begin{center}
Table 1: Summary of the results in the mirror assignment%

\begin{tabular}
[c]{|l|l|l|}\hline
Quantity & $N(1535)$ & $N(1650)$\\\hline
$m_{0}$ [MeV] & $459\pm117$ & $659\pm146$\\\hline
$a_{0,theor}^{(-)}$ [MeV$^{-1}$] & $(6.41\pm0.17)\cdot10^{-4}$ &
$(6.25\pm0.19)\cdot10^{-4}$\\\hline
$\Gamma_{N^{\ast}\rightarrow N\eta}^{theor}$[MeV] & $4.9\pm0.8$ & $12.5\pm
2.5$\\\hline
\end{tabular}

\bigskip

Table 2: Values from experiment, ChPT and Lattice%

\begin{tabular}
[c]{|l|l|}\hline
$a_{0,exp}^{(-)}$ [MeV$^{-1}$] & $(6.4\pm0.1)\cdot10^{-4}$\\\hline
$a_{0,exp}^{(+)}$ [MeV$^{-1}$] & $(-8.85\pm7.1)\cdot10^{-6}$\\\hline
$\Gamma_{N(1535)\rightarrow N\pi}^{exp}$ [MeV] & $67.5\pm11.2$\\\hline
$\Gamma_{N(1650)\rightarrow N\pi}^{exp}$ [MeV] & $105\pm21$\\\hline
$\Gamma_{N(1535)\rightarrow N\eta}^{exp}$[MeV] & $63.0\pm10.5$\\\hline
$\Gamma_{N(1650)\rightarrow N\eta}^{exp}$[MeV] & $15.0\pm3.0$\\\hline
$a_{0,ChPT}^{(-)}$ (from \cite{Baru:2010xn}) [MeV$^{-1}$] & $(6.16\pm
0.06)\cdot10^{-4}$\\\hline
$a_{0,ChPT}^{(+)}$ (from \cite{Baru:2010xn}) [MeV$^{-1}$] & $(5.44\pm
0.22)\cdot10^{-5}$\\\hline
$g_{A}^{N(1535)}$ (from \cite{Takahashi}) & $0.2\pm0.3$\\\hline
$g_{A}^{N(1650)}$ (from \cite{Takahashi}) & $0.55\pm0.2$\\\hline
\end{tabular}

\end{center}

In conclusion, the mirror assignment provides a good description of the
phenomenology. The only open issue for the choice $N(1535)$ is the
theoretically too small decay width $\Gamma_{N^{\ast}\rightarrow N\eta},$
while the choice $N(1650)$ is in good agreement with all the values. In the
future, one could try to combine both scenarios by including also the
positive-parity Roper state $N(1440)$ and studying a four-state mixing problem
\cite{Gallas:2009qp}. Another interesting possible source of an enhancement of
the decay of $N(1535)$ into $N\eta$ is the inclusion of terms involving the
chiral anomaly in the baryonic sector. Namely, our Lagrangian
(\ref{nucl lagra}) respects the $U(1)_{A}$ symmetry (the breaking of it is
confined to the mesonic sector). However, in order to limit the free
parameters of the model, a detailed study of the effect of the anomaly in the
baryonic sector can be undertaken only when baryons with strangeness are
included and thus much more experimental constraints can be put on the model.

\subsection{Naive assignment}

For $N^{\ast}\equiv N(1535)$ as the chiral partner of the nucleon, we
determine the four free parameters $\tilde{c}_{1}$, $\tilde{c}_{2}$,
$\tilde{c}_{12}$, and $Z$ by performing the very same fit done in the mirror
assignment, leading to the following results ($\chi_{\text{min}}^{2}=0.64$):
\begin{equation}
\tilde{c}_{1}=-2.3\pm0.1\;,\;\tilde{c}_{2}=7.1\pm2.7\;,\;\tilde{c}%
_{12}=-0.9\pm0.1\;,\text{ }Z=1.81\pm0.07\text{ .} \label{naiveparam}%
\end{equation}
The parameters $\tilde{\widehat{g}}_{1}$ and $\tilde{\widehat{g}}_{2}$ are
given by the masses of the nucleons and $\sigma_{0}=167$ MeV through eq.
(\ref{mN_ghut}): $\tilde{\widehat{g}}_{1}=5.6\pm0.6$ , $\tilde{\widehat{g}%
}_{2}=9.2\pm1.0$.

The result for the isospin-odd scattering length is (see Appendix A for the
analytical expression ):
\begin{equation}
a_{0}^{(-)}=(4.32\pm0.34)\cdot10^{-4}\;\mathrm{MeV}^{-1}\;.
\end{equation}
This value is \emph{not} in agreement with experimental result $a_{0,exp}%
^{(-)}=(6.4\pm0.1)\cdot10^{-4}\;\text{MeV}^{-1}$ \cite{gotta}. Namely, it is
off by several standard deviations from the experimental value. This mismatch
is an evident drawback of the naive assignment. (Namely, in this case the
influence of the heavier resonances does not improve, but worsens the results
of the lowest-order ChPT.)

Also in the naive assignment, the isospin-even scattering length $a_{0}^{(+)}$
depends on the mass of the scalar-isoscalar $\sigma$, see Appendix B for its
explicit form. The quantity $a_{0}^{(+)}$ is plotted as function of
$m_{\sigma}$ in Fig. 2, left panel. We observe that the value of $a_{0}^{(+)}$
reaches the experimental band for $m_{\sigma}=1000$-$1300$ MeV. At first sight
this result seems even in better agreement with the experiment than the mirror
assignment, because the favoured mass of $\sigma$ is approximately $1.3$ GeV
\cite{denis}. However, care is needed: As mentioned above, in the present
approach the lightest scalar state $f_{0}(500)$ is \emph{not} present (in
either assignment), although the latter resonance is known to couple strongly
to nucleons, see also the discussions in\ Refs. \cite{giuseppe,ml} and in the
conclusions. Therefore, at the present stage an agreement with data is
actually misleading.%

%TCIMACRO{\FRAME{ftbpFU}{4.3829in}{2.9369in}{0pt}{\Qcb{The isoscalar scattering
%length $a_{0}^{(+)}$ is plotted as a function of $m_{\sigma}$ in the mirror
%assignment (dashed curve for the partner $N(1535),$ dotted curve for the
%partner $N(1650)$) and in the naive assignment (solid line, which is valid for
%both partners $N(1535)$ and $N(1650)$). The experimental range is shown by the
%lower band. The chPT result of Ref. \cite{Baru:2010xn} is represented by the
%upper narrow band.}}{}{SiV.eps}{\special{ language "Scientific Word";
%type "GRAPHIC";  display "USEDEF";  valid_file "F";  width 4.3829in;
%height 2.9369in;  depth 0pt;  original-width 28.8138in;
%original-height 9.2206in;  cropleft "0";  croptop "1";  cropright "1";
%cropbottom "0";  filename 'siv.eps';file-properties "NPEU";}}}%
%BeginExpansion
\begin{figure}
[ptb]
\begin{center}
\includegraphics[
height=2.9369in,
width=4.3829in
]%
{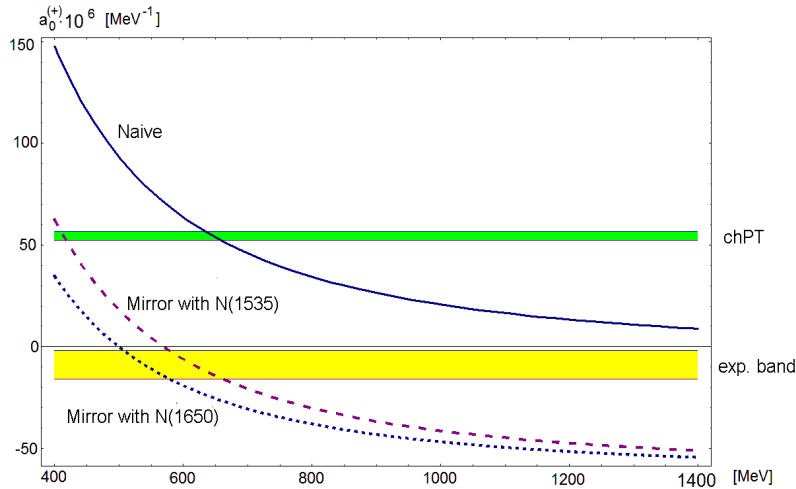}%
\caption{The isoscalar scattering length $a_{0}^{(+)}$ is plotted as a
function of $m_{\sigma}$ in the mirror assignment (dashed curve for the
partner $N(1535),$ dotted curve for the partner $N(1650)$) and in the naive
assignment (solid line, which is valid for both partners $N(1535)$ and
$N(1650)$). The experimental range is shown by the lower band. The chPT result
of Ref. \cite{Baru:2010xn} is represented by the upper narrow band.}%
\end{center}
\end{figure}
%EndExpansion

We now turn to the decay $N^{\ast}\rightarrow N\eta$, for which we obtain the
result
\begin{equation}
\Gamma_{N^{\ast}\rightarrow N\eta}=4.9\pm0.8\,\mathrm{MeV}\;,
\end{equation}
which is considerably smaller than the experimental value $\Gamma_{N^{\ast
}\rightarrow N\eta}^{\exp}=63.0\pm10.5$ MeV \cite{PDG}.

\begin{center}
Table 3: Summary of the results in the naive assignment%

\begin{tabular}
[c]{|l|l|l|}\hline
Quantity & $N(1535)$ & $N(1650)$\\\hline
$a_{0, theor}^{(-)}$ [MeV$^{-1}$] & $(4.32\pm0.34)\cdot10^{-4}$ &
$(4.31\pm0.34)\cdot10^{-4}$\\\hline
$\Gamma_{N^{\ast}\rightarrow N\eta}$[MeV] & $4.98\pm0.8$ & $12.5\pm
2.5$\\\hline
\end{tabular}

\end{center}

If we choose $N^{\ast}\equiv N(1650)$ as the parity partner of the nucleon in
the naive assignment, we get very small modifications of the scattering
lengths (see Table 2 for the isospin-odd one, Fig. 2 (right panel) for the
isospin-even one), but the decay width $\Gamma_{N^{\ast}\rightarrow N\eta
}^{\exp}=15\pm3$ MeV agrees with the experimental value just as in the mirror
assignment. We summarize the situation of the naive assignment in Table 3 and
Fig. 2. Note, in\ Fig. 2 only one curve for $a_{0}^{+}=a_{0}^{+}(m_{\sigma})$
is shown in the naive assignment: this is due to the fact that the results for
$a_{0}^{+}$ are almost identical for the two partners $N(1535)$ and $N(1650).$
The small numerical difference cannot be resolved graphically.

A last comment concerns the pion-nucleon sigma term in the naive assignment:
it amounts to $\sigma_{\pi N}\simeq18$ MeV, thus only slightly larger than in
the mirror case. A similar discussion to the one in Sec. 3.1 holds here.

\section{Discussions and conclusions}

\label{IV}

In this paper we have investigated the nucleon and its chiral partner,
identified with either $N(1535)$ or $N(1650)$, within the extended Linear
Sigma Model \cite{Gallas:2009qp,denis,Parganlija:2012fy}. The novel ingredient
is that the chiral partner was incorporated in the naive assignment in
presence of (axial-)vector degrees of freedom. The results were compared to
the ones obtained in \cite{Gallas:2009qp}, where the chiral partner was
coupled in the mirror assignment.

Two fundamental differences between the two assignments have played an
important role in this work:

\begin{itemize}
\item In the naive assignment, the mass generation takes place only through
the spontaneous breakdown of chiral symmetry. On the contrary, in the mirror
assignment the mass term
\[
\mathcal{L}_{m_{0}}=-m_{0}(\overline{\Psi}_{2}\gamma_{5}\Psi_{1}%
-\overline{\Psi}_{1}\gamma_{5}\Psi_{2})\;
\]
is allowed, and the mass of the nucleon is generated by this term \emph{and}
by the mechanism of chiral symmetry breaking. The quantity $\mathcal{L}%
_{m_{0}}$ is the only term in the baryonic sector of the eLSM Lagrangian which
does not fulfill dilatation invariance. In order to achieve this in a chirally
invariant way, which is one of the basic principle toward the construction of
the eLSM \cite{denis,Parganlija:2012fy}, one can modify $\mathcal{L}_{m_{0}}$
by including an interaction of the nucleon with a dilaton/glueball field $G$
and a tetraquark field $\chi$ (see Refs. \cite{scalars,tetraquark} and refs.
therein for a phenomenological discussion):
\begin{equation}
\mathcal{L}_{m_{0}}\rightarrow\mathcal{L}_{\chi G}=-(a\chi+bG)(\overline{\Psi
}_{2}\gamma_{5}\Psi_{1}-\overline{\Psi}_{1}\gamma_{5}\Psi_{2})\;.
\end{equation}
The mass term $m_{0}$ is then generated when shifting these fields by their
vacuum expectation values, $\chi\rightarrow\chi_{0}$ and $G\rightarrow G_{0}$:%
\begin{equation}
m_{0}=a\chi_{0}+bG_{0}\;\text{.}%
\end{equation}

This fact shows an important point: even if the scalar-isoscalar field
$\sigma$ (i.e., the chiral partner of the pion) is identified with the heavy
resonance $f_{0}(1370)$, as current results show
\cite{denis,Parganlija:2012fy,scalars,tetraquark}, in the mirror assignment it
is possible to write down a \emph{chirally invariant} interaction which
couples the nucleon to the lightest scalar-isoscalar light meson $f_{0}(500)$,
which can be identified with a tetraquark field $\chi$ (or, similarly, to a
pion-pion molecular state \cite{scalars,tetraquark}: the difference between
these two configurations is not important here). Moreover, one can couple the
nucleon also to the glueball field (to be identified predominantly with
$f_{0}(1500)$ or with $f_{0}(1710)$). These additional interactions are not
possible in the naive assignment without breaking chiral symmetry. This is a
clear theoretical advantage of the mirror assignment w.r.t. the naive
assignment. These properties are also important for the study of the model at
non-zero density: in Ref. \cite{giuseppe} it has been shown that a light
tetraquark field $\chi\equiv f_{0}(500)$ allows for a correct description of
nuclear matter saturation and compressibility.

\item In the naive assignment the interaction between $N$ and $N^{\ast}$ is
solely given by the derivative coupling to the pion, which in turn is possible
only due to the inclusion of the (axial-)vector mesons in the model. Without
(axial-)vector mesons the decay $N^{\ast}\rightarrow N\pi$ or the formation of
a resonance $N^{\ast}$ in $\pi N$ scattering does not take place. In the
mirror assignment we do not have such restrictions. Even without
(axial-)vector mesons the decay $N^{\ast}\rightarrow N\pi$ does not vanish
since terms of the form $\overline{N}i\gamma_{5}\vec{\pi}\cdot\vec{\tau
}N^{\ast}$ are present.
\end{itemize}

In the mirror assignment, the theoretical value of the isospin-odd scattering
length $a_{0}^{(-)}$ is in very well agreement with experimentally measured
and the ChPT results (see Tables 1 and 2). This is not the case in the naive
assignment (Table 3), in which the theoretical result $a_{0}^{(-)}$ is clearly
too small. The isospin-even scattering is not yet conclusive (see Fig. 2): one
has to go beyond the present treatment by including both resonances
$f_{0}(500)$ and $f_{0}(1370)$ in the study of the scattering lengths, see
also Refs. \cite{giuseppe,ml}. The decay into $N\eta$ turns out to bee too
small for $N(1535)$ in any assignment, whereas for $N(1650)$ it agrees with
the experimental data, thus favouring the latter as the chiral partner of the
nucleon. Summarizing, our outcome is the following: the mirror assignment with
the resonance $N(1650)$ as the chiral partner of the nucleon seems to be the
favoured situation. Future studies on baryons are necessary to confirm that
the mirror assignment is the correct way to incorporate the nucleon and its
chiral partner. In the vacuum, an important future work is the description in
a unique framework of the four baryonic resonances $N(940),$ $N(1440),$
$N(1535),$ and $N(1650)$ \cite{Gallas:2009qp}. In this enlarged scenario it
will be possible to test further the mirror and the naive assignments (and
eventually a combination of them). Important hints come also from studies at
nonzero density \cite{giuseppe,ziesche}, where it is shown that the mirror
assignment is able to describe nuclear matter at nonzero density; the results
do not depend on the mass of the chiral partner (and thus on the choice of
$N(1535)$ or $N(1650)$), but depend strongly on the value of $m_{0}$. Namely,
only for $m_{0}\gtrsim500$ MeV the nuclear matter compressibility can be
correctly described \cite{giuseppe}. The range $m_{0}\gtrsim500$ MeV is
compatible with $N(1535)$, for which we have $m_{0}=459\pm117$ MeV, but also
(and even better) with $N(1650),$ for which $m_{0}=659\pm146$ MeV. The need of
a sizable $m_{0}$ was also corroborated in the recent study of Ref.
\cite{achinnew}, in which the emergence of an inhomogenous chiral condensate
was studied: it was shown that a value of $m_{0}\gtrsim500$ MeV is necessary
to have an homogeneous liquid nuclear matter ground state.

In this paper we have also evaluated the pion-nucleon sigma term, which turns
out to be about $15$ MeV in the mirror case and about $18$ MeV in the naive
one. Both values are smaller than the results present in the literature. This
is due to the fact that in our approach no explicit symmetry breaking is
considered in the baryonic part of the Lagrangian(s) and the pion-nucleon
sigma terms arise solely from the explicit breaking present in the chiral
condensate. The result represents anyhow a sizable contribution to the
pion-nucleon sigma term. The missing piece can be included by adding an
explicit breaking mass term for the baryons. This is an important outlook,
which is also necessary for a better description of the isospin-even
scattering length.

Considering that the calculation of scattering lengths and decays was carried
at tree-level, an interesting outlook is the inclusion of mesonic loops.
However, many resonances are present in the model, which are exchanged in
scattering processes (as, for instance, the (axial-)vector mesons); their
contribution is automatically taken into account at tree-level without the
need of resummations. Moreover, the ratio of the decay width to the mass of
the $N^{\ast}$ resonance (in both scenarios) is small, implying that the
evaluation of loops, while surely valuable, is not expected to change the
qualitative picture that we have presented \cite{loop}.

In future studies one should also incorporate the $\Delta$ resonance
\cite{ellis,jidoprl} and extend the present model to $N_{f}=3$ also in the
baryonic sector.

\bigskip

\textbf{Acknowledgments:} The authors thank D. H. Rischke, G. Pagliara, K.
Teilab, C. Sasaki, M. Rho, A. Heinz, H. K. Lee, and J. Schaffner-Bielich for
useful discussions. S. G. acknowledges support by the BMBF (Projekt
05P09RFFTF) and F. G. by the `Stiftung Polytechnische Gesellschaft - Frankfurt
am Main'.

\appendix

\section{Low-energy phenomenology}

\label{appendixA}

In this appendix we present the formulae for the isospin-even and isospin-odd
$s$-wave scattering lengths, for the axial coupling constant of the nucleon
and its chiral partner, and for the decay widths $\Gamma_{N^{\ast}\rightarrow
N\pi}$ and $\Gamma_{N^{\ast}\rightarrow N\eta}$. It should be stressed that
the results presented here are based on a tree-level calculation.

The isospin-even scattering length takes the form:%

\begin{align}
a_{0}^{(+)} &  =\frac{1}{4\pi(1+\frac{m_{\pi}}{m_{N}})}\left\{  -2Z^{2}%
w^{2}\tilde{c}_{12}^{2}(m_{N^{\ast}}-m_{N})\left[  1+\frac{(m_{N^{\ast}}%
^{2}-m_{N}^{2})(m_{N}^{2}+m_{\pi}^{2}-m_{N^{\ast}}^{2})}{(m_{N}^{2}+m_{\pi
}^{2}-m_{N^{\ast}}^{2})^{2}-4m_{N}^{2}m_{\pi}^{2}}\right]  \right.
\nonumber\label{isoskalarnaive}\\
&  -4Z^{2}w\tilde{c}_{1}(\tilde{\widehat{g}}_{1}-\tilde{c}_{1}wm_{N}%
)+\frac{\tilde{\widehat{g}}_{1}}{m_{\sigma}^{2}}\left[  2g_{1}wm_{\pi}%
^{2}+\frac{Z}{f_{\pi}}\left(  m_{\sigma}^{2}-\frac{m_{\pi}^{2}}{Z^{2}}\right)
-2\frac{Zw^{2}m_{\pi}^{2}m_{1}^{2}}{f_{\pi}}\right]  \nonumber\\
&  +4m_{\pi}m_{N}Z^{2}\left[  w^{2}\tilde{c}_{12}^{2}\frac{m_{\pi}(m_{N^{\ast
}}-m_{N})^{2}}{(m_{N}^{2}+m_{\pi}^{2}-m_{N^{\ast}}^{2})^{2}-4m_{N}^{2}m_{\pi
}^{2}}-\left.  \frac{(\tilde{\widehat{g}}_{1}-2\tilde{c}_{1}wm_{N})^{2}%
}{4m_{\pi}(m_{N}^{2}-m_{\pi}^{2})}\right]  \right\}  \;.
\end{align}
The isospin-odd scattering length is given by:
\begin{align}
a_{0}^{(-)} &  =\frac{1}{4\pi(1+\frac{m_{\pi}}{m_{N}})}\left\{  4Z^{2}%
w^{2}\tilde{c}_{12}^{2}\frac{m_{N}m_{\pi}(m_{N^{\ast}}-m_{N})^{2}(m_{N^{\ast}%
}+m_{N})}{(m_{N}^{2}+m_{\pi}^{2}-m_{N^{\ast}}^{2})^{2}-4m_{N}^{2}m_{\pi}^{2}%
}\right.  \nonumber\label{isovectornaive}\\
&  -m_{\pi}2Z^{2}w^{2}\tilde{c}_{12}^{2}\left[  1+\frac{(m_{N^{\ast}}%
-m_{N})^{2}(m_{N}^{2}+m_{\pi}^{2}-m_{N^{\ast}}^{2})}{(m_{N}^{2}+m_{\pi}%
^{2}-m_{N^{\ast}}^{2})^{2}-4m_{N}^{2}m_{\pi}^{2}}\right]  \nonumber\\
&  \left.  -2Z^{2}\frac{(\tilde{\widehat{g}}_{1}-2\tilde{c}_{1}wm_{N})^{2}%
}{4m_{N}^{2}-m_{\pi}^{2}}-2Z^{2}w^{2}\tilde{c}_{1}^{2}+\frac{2w\tilde{c}_{1}%
}{Zf_{\pi}}\left(  \frac{m_{a}^{2}}{m_{\rho}^{2}}\right)  \right\}  .
\end{align}
We have verified that the scattering lengths vanish in the chiral limit,
$m_{\pi}\rightarrow0,$ as required by the low-energy theorems of QCD.

The expression for the decay width $N^{\ast}\rightarrow N\pi$ is:
\begin{equation}
\Gamma_{N^{\ast}\rightarrow N\pi}=3\,\frac{k_{\pi}}{2\pi}\,\frac{m_{N}%
}{m_{N^{\ast}}}\,\frac{Z^{2}w^{2}\tilde{c}_{12}^{2}}{2}\left[  (m_{N^{\ast}%
}^{2}-m_{N}^{2}-m_{\pi}^{2})\,\frac{E_{\pi}}{m_{N}}+m_{\pi}^{2}\,\left(
1-\frac{E_{N}}{m_{N}}\right)  \right]  \label{naiverZerfall}%
\end{equation}
where the momentum of the pseudoscalar particle is given by
\begin{equation}
k_{\pi}=\frac{1}{2m_{N^{\ast}}}\sqrt{(m_{N^{\ast}}^{2}-m_{N}^{2}-m_{\pi}%
^{2})^{2}-4\,m_{N}^{2}m_{\pi}^{2}}\;\text{,} \label{kpion}%
\end{equation}
and where the energies are $E_{\pi}=\sqrt{k_{\pi}^{2}+m_{\pi}^{2}}$ and
$E_{N}=\sqrt{k_{\pi}^{2}+m_{N}^{2}}$. Note, $\Gamma_{N^{\ast}\rightarrow N\pi
}$ vanishes in the limit in which the (axial-)vector meson decouple,
$g_{1}=0\rightarrow w=0.$

Finally, the expression for the decay width $N^{\ast}\rightarrow N\eta$ takes
the form%
\begin{equation}
\Gamma_{N^{\ast}\rightarrow N\eta}=\text{cos}^{2}\phi\,\frac{k_{\eta}}{2\pi
}\,\frac{m_{N}}{m_{N^{\ast}}}\,\frac{Z^{2}w^{2}\tilde{c}_{12}^{2}}{2}\left[
(m_{N^{\ast}}^{2}-m_{N}^{2}-m_{\eta}^{2})\,\frac{E_{\eta}}{m_{N}}+m_{\eta}%
^{2}\,\left(  1-\frac{E_{N}}{m_{N}}\right)  \right]
\end{equation}
where $k_{\eta}$ and $E_{\eta}$ are the momentum and the energy of the $\eta
$-meson. The factor $\text{cos}^{2}\phi$ comes from the fact that the flavour
wave function of the $\eta$-meson is given by
\begin{equation}
\eta=\eta_{N}\cos\phi+\eta_{S}\sin\phi\;,
\end{equation}
with $\eta_{N}\equiv(\overline{u}u+\overline{d}d)/\sqrt{2}$ and $\eta
_{S}\equiv\overline{s}s$. Hence, the decay amplitude $\mathcal{A}_{N^{\ast
}\rightarrow N\eta}$ should be written as:
\begin{equation}
\mathcal{A}_{N^{\ast}\rightarrow N\eta}=\mathcal{A}_{N^{\ast}\rightarrow
N\eta_{N}}\,\cos\phi+\mathcal{A}_{N^{\ast}\rightarrow N\eta_{S}}\,\sin\phi\;.
\end{equation}
We neglect $\mathcal{A}_{N^{\ast}\rightarrow N\eta_{S}}$ because it is
OZI-suppressed and we use the mixing angle $\phi=-40^{\circ},$ in the middle
of the range given in Eq. \cite{Venugopal:1998fq}

\section{Naive model without (axial-)vector mesons: decoupling of $N$ and
$N^{\ast}$}

\label{appendixB}

The \textit{naive} coupling constants of Eq. (\ref{Lagnaive}) are linked to
those of Eq. (\ref{Lagnaive2}) in the following way:
\begin{align}
\tilde{\widehat{g}}_{1}  &  =\frac{-1}{4\;\mathrm{cosh}\delta}\left(
2\widehat{g}_{12}+\widehat{g}_{1}e^{\delta}-\widehat{g}_{2}e^{-\delta}\right)
,\nonumber\\
\tilde{\widehat{g}}_{2}  &  =\frac{-1}{4\;\mathrm{cosh}\delta}\left(
2\widehat{g}_{12}-\widehat{g}_{1}e^{-\delta}+\widehat{g}_{2}e^{\delta}\right)
,\nonumber\\
\tilde{c}_{1}  &  =\frac{1}{4\;\mathrm{cosh}\delta}\left(  2c_{12}%
+c_{1}e^{\delta}+c_{2}e^{-\delta}\right)  ,\nonumber\\
\tilde{c}_{2}  &  =\frac{1}{4\;\mathrm{cosh}\delta}\left(  -2c_{12}%
+c_{1}e^{-\delta}+c_{2}e^{\delta}\right)  ,\nonumber\\
\tilde{c}_{12}  &  =\frac{1}{4\;\mathrm{cosh}\delta}\left(  -2c_{12}%
\;\mathrm{sinh}\;\delta+c_{1}-c_{2}\right)  . \label{geschlaengelt}%
\end{align}

As a last step we show explicitly the decoupling of the nucleon and the chiral
partner in the naive assignment without (axial-)vector mesons. In this case
the Lagrangian (\ref{Lagnaive}) takes the form:
\begin{align}
\mathcal{L}_{naive}  &  =\overline{\Psi}_{1}i\gamma_{\mu}\partial^{\mu}%
\Psi_{1}-\widehat{g}_{1}\overline{\Psi}_{1}(\sigma+i\gamma_{5}\vec{\tau}%
\cdot\vec{\pi})\Psi_{1}+\overline{\Psi}_{2}i\gamma_{\mu}\partial^{\mu}\Psi
_{2}-\widehat{g}_{2}\overline{\Psi}_{2}(\sigma+i\gamma_{5}\vec{\tau}\cdot
\vec{\pi})\Psi_{2}\nonumber\\
&  -\widehat{g}_{12}\overline{\Psi}_{1}(\gamma_{5}\sigma+i\vec{\tau}\cdot
\vec{\pi})\Psi_{2}+\widehat{g}_{12}\overline{\Psi}_{2}(\gamma_{5}\sigma
+i\vec{\tau}\cdot\vec{\pi})\Psi_{1}+...
\end{align}
where we have set $c_{1}=c_{2}=c_{12}=g_{1}=w=0$ and where, for simplicity,
the scalar mesons as $\eta$ and $\vec{a}_{0}$ are omitted. After introducing
the physical fields $N$ and $N^{\ast}$ as in eq. (\ref{rotationfelder}) we
get:
\[
\mathcal{L}=\overline{N}i\gamma_{\mu}\partial^{\mu}N-\tilde{\widehat{g}}%
_{1}\overline{N}(\sigma+i\vec{\tau}\cdot\vec{\pi})N+\overline{N}^{\ast}%
i\gamma_{\mu}\partial^{\mu}N^{\ast}-\tilde{\widehat{g}}_{1}\overline{N}^{\ast
}(\sigma+i\vec{\tau}\cdot\vec{\pi})N^{\ast},
\]
Thus, one has a full decoupling of the fields $N$ and $N^{\ast}$: there is no
interaction among them \cite{Jido:2001nt}. Only by introducing the
(axial-)vector mesons the desired interactions arise.

\end{document}